\input stromlo.sty

\title Structure, Formation and Ages of Elliptical Galaxies

\shorttitle Structure, Formation and Ages of Ellipticals 

\author Ralf Bender

\shortauthor Ralf Bender

\affil Universit\"ats-Sternwarte, Scheinerstr. 1, D--81679 M\"unchen, Germany

\abstract 

The structural properties of elliptical galaxies are consistent with
their formation in a merging hierarchy. In this picture, the role of
gaseous processes and dissipation decreased with increasing mass
creating preferentially rotationally flattend disky ellipticals (and
S0s) at lower luminosities and boxy, anisotropic ellipticals (often with
peculiar kinematics) at higher luminosities. However, gas
and dissipation processes must have been important even in the
formation of the most luminous ellipticals. They played key roles in
determining the phase space structure of Es and possibly ensured a tight
fundamental plane.

The bulk of the stars in the majority of {\it luminous} cluster
ellipticals formed at redshifts above two, likely above three. This
follows from their homogenous colors and line strengths and their
essentially passive evolution with redshift.  Low luminosity
ellipticals (and probably field ellipticals) may have had extended
star formation histories, possibly associated with the presence of
disks. The star-formation time-scales can presumably be constrained on
the basis of abundance ratios and seem to be {\it inversely} related
to galaxy mass. Massive ellipticals are likely to have formed the bulk
of their stars within one Gigayear.

\section Introduction

Elliptical Galaxies can appear as complex individuals when we analyse
their detailed structure and internal dynamics. But they can also
behave highly uniformly when viewed under the perspective of global
scaling properties or stellar populations. As will be discussed in
this paper, these apparently disparate properties are keys to their
nature and allow to put constraints on their formation histories and
ages.

In Section 2, I review the varieties of structures we observe in
ellipticals and analyse the relevance of dissipation and merging in
their formation. In Section 3, I discuss a few aspects of the
fundamental plane. In Sections 4 and 5, constraints on the ages and
star formation time scales of ellipticals are derived on the basis of
their stellar populations. In Section 6, I set out to higher redshift
to check whether the inferences derived from the local population of
ellipticals find confirmation in the evolution of their properties.

\section Structure and Formation Processes

There is convincing evidence that Es can be sub-devided into
basically two groups with respect to their structural properties. One
group is characterized by boxiness, anisotropy and shallow cores, the
other by diskiness, rotational flattening and the absence of cores
(e.g. Bender 1988a, 1990, Faber et al. 1997). It may well be that there
exists a clear-cut separation between the two groups, but this is not
clear yet.

Disky Es contain faint disks which contribute between a few
percent and up to 30\% to the total light of the galaxy (Rix \& White
1990, Scorza \& Bender 1995). Generally, the spheroids are
rotationally supported and the angular momenta of disks and bulges are
parallel to each other indicating a 'coordinated' formation (Bender et
al. 1993). Recent
HST imaging by Faber et al. (1997) has shown that disky Es also have
high density power-law profiles that lack cores while boxy Es
have shallow cores (confirming and extending earlier claims by Nieto et
al. 1991). These data show that dissipation was essential for the
formation of disky Es. They simply seem to form the
continuation of the Hubble sequence to the lowest disk-to-bulge ratios
(Bender 1988a, 1990, Kormendy \& Bender 1996).

Boxy Es, on the other hand, are mostly supported by
anisotropic velocity dispersions and frequently show hints for a
formation dominated by merging processes (Nieto 1988, 
Bender \& Surma 1992). Peculiar velocity fields, like minor axis
rotation and kinematically decoupled cores (Franx \& Illingworth 1988,
Jedrezjewski \& Schechter 1988, Bender 1988b) are a natural by-product of
merging of star-dominated systems (Hernquist \& Barnes 1991;
in a few cases they could also have
other origins, see Statler 1994). Unlike shells or ripples
(e.g., Schweizer 1990), these features are long-lived or permanent and
carry 'genetic' information about the formation process of the main
(i.e. inner) parts of the galaxy. Interestingly, disky and boxy
Es are also separated by their radio and X-ray properties
(Bender et al. 1989). This indicates interesting links between galaxy
structure and the presence or feeding of black holes as well as the
depth of the galaxies' potential wells.

The dynamical structure and the degree of peculiarity of the merger
product are presumably strongly related to the gas content of the
progenitors. Little gas seems to suffice to reduce the fraction of box
orbits in the merger remnant in favor of z-axis tubes (Barnes \& Hernquist 
1996). Centrally concentrated gas, like central cusps, may also
destabilize box orbits (Dubinski 1994). Since box orbits are the
backbone of triaxial objects (see e.g. Merritt 1997), while z-tubes
are dominant in rotationally flattened objects, the gas fraction in the
progenitors must be a key factor in determining the structure of the
merger remnant. Simply speaking, less gas implies a more exciting end
product.  In this sense, we cannot exclude that disky Es are
merger products as well. We can only conclude that, if they are, gas
has dominated sufficiently to remove all clear-cut evidence for their
merger history.

In a hierarchical galaxy formation scenario galaxies are expected to
form via a sequence of merging and accretion processes (e.g. White
1995).  Presumably, merging is also driving the star formation history
(e.g. Katz 1992, Steinmetz \& M\"uller 1994) depleting the gas
with increasing galaxy mass. Therefore, the formation of the most
massive galaxies involved rather little gas and dissipation. 
Consistent with this picture,
the vast majority of boxy Es has luminosities above
$L_*$, while the luminosity function of disky Es resembles the one of
S0s (Bender et al. 1993).

The low-redshift analogue of the late formation-phase of boxy Es may
be found in ultraluminous IRAS mergers (Schweizer 1990, Kormendy \&
Sanders 1992, Bender \& Surma 1992). While violent relaxation in these
objects will likely create a boxy main body (Steinmetz 1995), the
molecular gas concentrates in the central kiloparsec and can form a
kinematically decoupled core (Hernquist \& Barnes 1991, Barnes 1996). 
Indeed, the
masses and metallicities of kinematically decoupled cores in Es are
quite similar to those of the central gas tori in IRAS mergers (Bender
\& Surma 1992).  We can speculate that very gas-rich progenitors may
also create disky Es. The analogy of E galaxy formation and
the IRAS merging process is unlikely to be perfect.  Especially, it
does not necessarily imply that Es generally formed in
spiral-spiral mergers at low redshift -- merging of any star-dominated
progenitors at any redshift may have produced similar remnants.

\section Fundamental Plane

Ellipticals define a two-dimensional manifold in the three-dimensional
space of their global structural parameters (effective radius $R_e$,
mean effective surface brightness $<SB>_e$, velocity dispersion
$\sigma$), the so-called fundamental plane (Djorgovski
\& Davis 1987, Dressler et al. 1987). Its defining relation is 
$\log R_e = 1.25 \log\sigma + 0.32 <{\rm SB}>_e + const.$ (e.g. J\o
rgensen et al. 1996). It seems to be independent from environment (J\o
rgensen et al. 1996) and is also valid for S0s and, with slight
changes, for dwarf ellipticals, too (Bender, Burstein \& Faber
1992). It is now generally agreed that the fundamental plane is simply
a consequence of the Virial theorem and the fact that E galaxies have
similar mass-to-light ratios and close to homologous structure at a
given luminosity (e.g. Faber et al. 1987, Djorgovski et al. 1989,
Bender, Burstein \& Faber 1992). There are two aspects of the
fundamental plane. Roughly speaking, the edge-on view is indicative of
the degree of similarity between Es of a given luminosity, the face-on
view provides information about formation processes and evolution.

Despite of the large variety of internal dynamics and structure, the
scatter perpendicular to the fundamental plane is very
small. Jorgensen et al. (1996) find a typical rms-scatter of 20\% in
$R_e$. In the case of the Coma cluster, the scatter is smaller than
10\% (Saglia et al. 1993, 1997), which is quite surprising for such
complex objects as Es (see Figure 1 below). One explanation
for this regularity could be the presence of at least some gas in all
merging events Es underwent (see above). If the gas fraction
was always about the same at a given mass then Es of similar
luminosity may have similar phase space structure despite of peculiar
velocity fields (which may be just the frost on the cake --- however
frost that carries the finger-prints of formation...). This notion is
also supported by kinematic and photometric studies which suggest that
ellipticals are only mildly triaxial and generally close to oblate
(e.g. de Zeeuw \& Franx 1991). This is significantly different from the
prolate-triaxial shape of, e.g., dark matter halos formed in
collisionless collapse (e.g. Frenk et al. 1988).

Renzini \& Ciotti (1993), Ciotti et al. (1996), Graham \& Colless
(1997) and others have carried out detailed studies about the
influence of various parameters on the scatter perpendicular to the
fundamental plane.  They derive interesting constraints on the ages,
the initial mass function of the stars, the variation of dynamical
structure and density distribution and on dark matter content.  The
implied small scatter in mass-to-light ratio constrains the variation
of ages at a given luminosity significantly and is consistent with
other indications based on colors and line-strengths (see next
section). Graham \& Colless (1997) find convincing evidence that the
tilt of the fundamental plane relative to the simple virial relation is
due to the systematic variation of surface brightness with luminosity.

Bender, Burstein \& Faber (1992) have analysed the distribution of Es
and other galaxies {\it within} the fundamental plane. The face-on
view of the fundamental plane is very similar to the
luminosity-surface-brightness diagram of Es. Together with bulges, E
galaxies define a sequence in the plane that extends from high
density, low luminosity objects like M32 to low density, high
luminosity objects like M87.  Several properties vary smoothly with
mass along this sequence (though with significant scatter), including
bulge-to-disk ratio, radio and X-ray properties, rotation, degree of
velocity anisotropy, and peculiar kinematics.  These trends are
consistent with the idea that the final mergers leading to larger
galaxies were systematically more stellar (and less gaseous) than
those producing smaller galaxies, as discussed above.  If compared
with typical cold-dark-matter fluctuations, it also follows that the
baryonic component in high luminosity Es dissipated less than the one
of low luminosity Es.

\section Stellar Populations and Ages

The stellar populations of elliptical galaxies are surprisingly
homogenous, consistent with the small scatter about the fundamental plane.
Colors and line-strengths are generally one-to-one correlated and
scale with luminosity or, even more tightly, with velocity dispersion
$\sigma$ (e.g.  Burstein et al. 1988, Dressler et al. 1987, Bower et
al. 1992, Bender, Burstein \& Faber 1993). It is important to note
that there is no difference in the Mg$-\sigma$ relation between disky
and boxy Es or between kinematically peculiar and regular Es.
However, there are hints for a weak dependence of the
color$-\sigma$ and Mg$-\sigma$ relations on the presence of rather
short-lived peculiarities (due to accretion of younger stars,
Schweizer et al. 1990) and, possibly, on environmental density (Lucey
et al. 1993, J\o rgensen 1997).

With stellar population synthesis models (e.g., Worthey 1994)
one can estimate the combined scatter in age and metallicity from the
observed scatter in the color$-\sigma$ or Mg$-\sigma$ relations. 
Consistently, Bower et al. (1992) and Bender, Burstein \& Faber
(1993) found for {\it luminous cluster} Es that the scatter
in age and/or metallicity at a fixed $\sigma$ must be smaller than
15\%. Evidently, luminous Es (independent of whether they are
boxy or not) cannot have formed continously over the Hubble time
suggesting that they are old on average.  High age and small
metallicity spread are also required to explain the observed very high
Mg-absorption in the central parts of Es (Greggio 1997).  

The age constraints for field Es and lower luminosity Es,
which mostly belong to the disky class, are
much less tight due to small samples or larger scatter in Mg and colors 
at smaller $\sigma$.  In
fact, it is indicated that low-luminosity Es ($\rm M_T \approx
-18$) seem to be systematically younger than giant Es ($\rm M_T
\approx -21$), see Faber et al. (1995) and Worthey (1996). Note that
this trend runs opposite to the one expected in a cold-dark-matter
model (Kauffmann et al. 1994). The apparently smaller ages of low
luminosity Es could in fact be caused by the faint disks they
contain. These disks may become more dominant towards lower
luminosities and may have had extended star formation histories. Hints
for this have been found by de Jong \& Davies (1996).

\section Abundance Ratios and Star Formation Time Scales

Another way to extract information about the star formation history of
ellipticals is to analyse their element abundance ratios.  For {\it
luminous} Es, Worthey et al. (1992), Davies et al. (1993) and others
found consistently that Mg is overabundant relative to Fe. Over a
larger luminosity range, [Mg/Fe] seems to be correlated with velocity
dispersion: faint Es have [Mg/Fe]$ \approx 0$, while luminous Es reach
[Mg/Fe]$\approx 0.4$ (Gonzalez 1993, Fisher, Franx \& Illingworth
1995). Furthermore, Paquet (1994) could show that, in luminous Es,
other light elements like Na and CN are overabundant relative to Fe as
well. Within the galaxies, the [Mg/Fe] overabundance is usually
radially constant up to at least the effective radii (Davies et
al. 1993, Paquet 1994). Generally, no distinction between 'normal'
luminous Es and Es with kinematically decoupled cores is indicated.
This implies that the enrichment history of luminous Es differed
significantly from the one of the solar neighborhood, see e.g.,
Matteucci \& Greggio (1986), Truran \& Burkert (1995), Faber et
al. (1995), Worthey (1996).

Evidently, the enrichment of {\it massive} (high velocity dispersion)
Es was dominated by Supernovae II, as only they can produce a
light element overabundance.  Supernovae Ia basically just provide
iron peak elements (see, e.g., Truran \& Thielemann 1986). Because
the yields of SNII integrated over a plausible IMF result in
[Mg/Fe]$\approx 0.3\,$dex at most (see Thomas, Greggio \& Bender 1997
who use the most up-to-date yield estimates), we
can conclude that the contribution of SNIa to the enrichment of the
most massive Es must have been small, if not negligible.

The prevalence of Supernovae II and in turn the light element
overabundance in massive Es can have the following reasons:
(a) a star formation time scale smaller than about 1Gyr (SNI explode
in significant numbers only after a few times 10$^8$yrs after star
formation started, e.g. Truran \& Burkert 1995), (b) a top heavy
initial mass function, (c) a reduced frequency of binary stars
(leading to fewer SNI events). -- Option (c) is
rather unlikely because one expects the binary frequency to be
determined by the local process of star formation rather than by
global galaxy properties. In addition, a low binary fraction may be
inconsistent with the observed frequency of discrete X-ray sources in
old populations.  Neither does option (b) work well, because the
overabundance in massive Es reaches [Mg/Fe]$\approx 0.4\,$dex
(as is also observed in Galactic halo stars, Fuhrmann et
al. 1995). For such high overabundances, a flat IMF alone cannot solve
the overabundance problem. So, option (a), i.e. a short star 
formation time scale, seems to be necessary in any case.
In their recent study, Thomas, Greggio \& Bender (1997) show
that the star formation time scale in massive Es was probably
shorter than roughly 1Gyr.

Note that these considerations do not only apply to the cores of
luminous Es but for the bulk of their stars, since the [Mg/Fe]
overabundane is similar at all radii (see above).  And another
important conclusion can be drawn from these findings: since most
present day spirals have gas-to-star ratios smaller than 0.2 and disk
stars show solar element ratios, merging of objects similar to
present-day spirals cannot produce objects similar to most present-day
massive Es.  However, even some luminous Es (e.g. NGC 5322) have
[Mg/Fe]$\approx 0$ and could be late merger products.

Since lower luminosity ellipticals have smaller light-element
overabundances, their star formation time scales are not severely
constrained. In fact, solar element abundance ratios could be
taken as a hint for extended star formation histories in smaller Es.

\section Redshift Evolution

The amount of data on luminous elliptical galaxies at intermediate and
high redshifts is now rapidly increasing thanks to bigger telescopes
and better instruments. This allows to study the evolution of their
luminosities (Glazebrook et al. 1995, Lilly et al. 1996), colors
(Aragon--Salamanca et al. 1993, Stanford et al. 1995) and surface
brightnesses (Dickinson 1995, Pahre et al. 1996, Schade et al. 1996).
The tightest constraints are derived on the basis of the Mg$-\sigma$
relation (Bender, Ziegler \& Bruzual 1996) and the fundamental plane
(Franx 1993, 1995, van Dokkum \& Franx 1996, Bender et al. 1997). All
data indicate that the redshift evolution of massive Es is very small
and basically consistent with passive evolution of very old stellar
populations, there is no evidence for dynamical evolution. The bulk of
the stars in massive Es must have formed at redshifts $z>2$, likely at
$z>3$.

\figureps[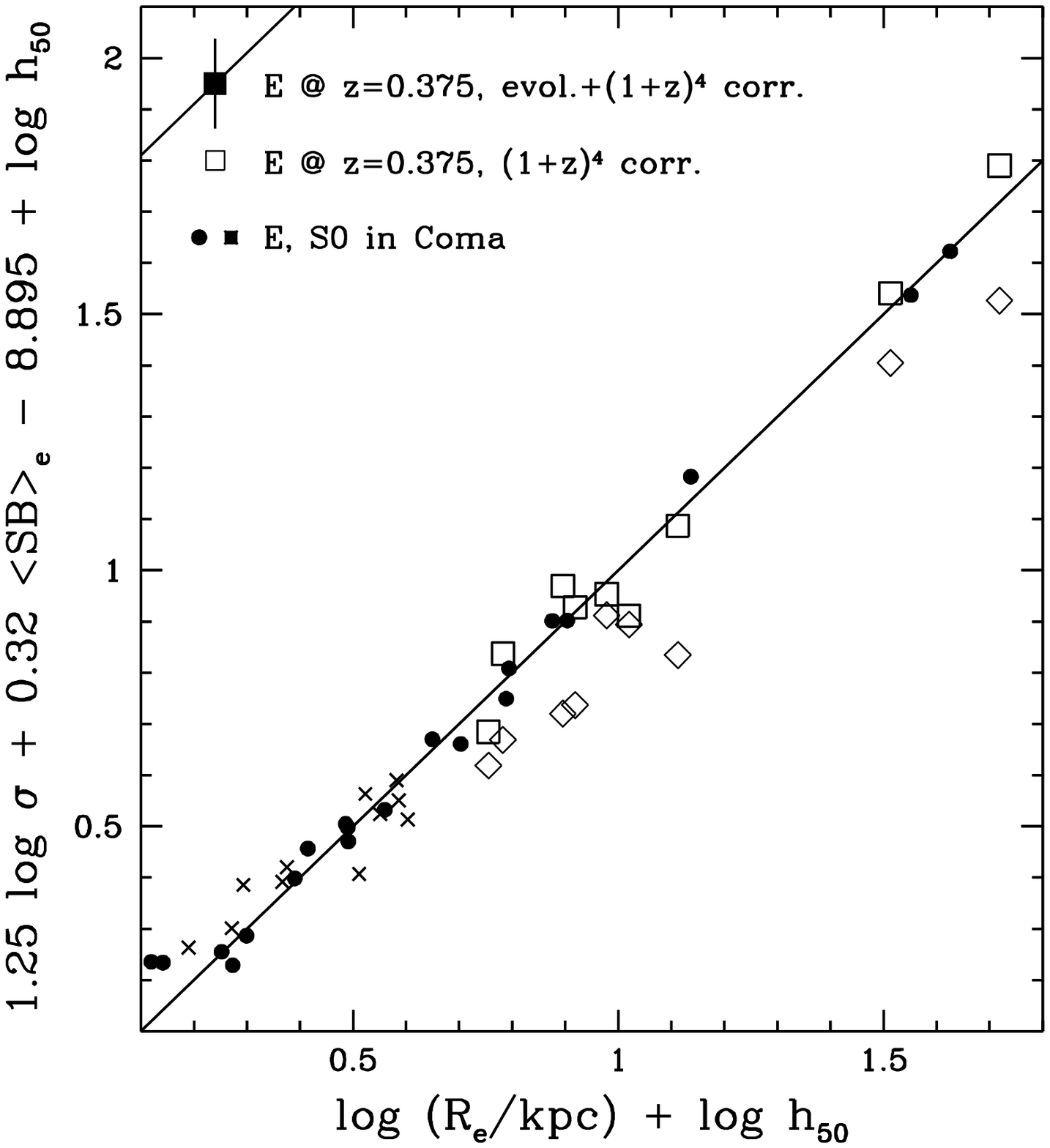,.55\hsize] 1. The Fundamental Plane of
E and S0 galaxies. Small filled circles and small crosses
denote E  and S0 galaxies, respectively, in the Coma cluster with
velocity dispersions \hbox{$\sigma > 120~$km/s} and effective radii
$R_e$ and effective surface brightnesses $\langle$SB$\rangle_e$ in the
B band.  The typical measurement errors are somewhat smaller than the
scatter indicates.  The open diamonds show Es at $z = 0.375$, also in
rest-frame B-band and after surface brightness has been corrected for
cosmological $(1+z)^4$ dimming. The open squares represent the $z = 0.375$
Es after further correction for luminosity evolution as derived from
the Mg$_b-\sigma$ relation (see text).  
A typical error bar is shown for the $z =
0.375$ Es in the upper left. h$_{50}$ is the Hubble constant in units
of 50~km/s/Mpc.  The Figure is adapted from Bender et al. (1997).

As an example, Figure 1 shows the fundamental plane of Es in the
clusters Abell 370 and MS1512+36 at $z=0.375$. Details about the
observational procedure can be found in Ziegler \& Bender 1997 and
Saglia et al. (1997).  The surface brightness term has been
transformed to rest--frame B--band and corrected for cosmological
dimming ($(1+z)^4$). The difference between the fundamental plane at
$z=0.375$ and in Coma is very small and is entirely accounted for by
passive evolution. Passive evolution of the stellar population can be
corrected for using the reduced Mg-absorption of distant Es as derived
from the ${\rm Mg}_b-\sigma$ relation.  From Worthey's (1994) models
one obtains the relation between Mg$_b$ and B-band evolution: $\Delta
B = 1.4\Delta{\rm Mg}_b$ (Bender et al. 1997).  Once this correction is
applied, the Es at $z=0.375$ fall on top of the Es in the Coma
cluster.

It is important to note that these results on the redshift evolution
of Es, and especially the most reliable ones, refer mostly to luminous
Es or Es in clusters and so they may not apply to low luminosity Es or
field Es. Indeed, there is a hint that at least in the Lilly et
al. (1996) sample, which contains mostly bright field Es, the number
density of Es evolves with redshift (Kauffmann et al. 1997). So, a
sizable fraction of field Es could in principle still form at
lower redshifts.

The evolution of Es in clusters cannot be discussed without
considering the fate of blue cluster members and E$+$A galaxies in
intermediate redshift clusters (Butcher \& Oemler 1978, Dressler \&
Gunn 1983). As recent investigations indicate, these objects are
unlikely to end as {massive} Es.  HST imaging shows that most blue
cluster galaxies and a significant fraction of E$+$A galaxies are in
fact infalling spirals or irregular galaxies (Dressler et al. 1994,
Wirth et al. 1994, Belloni et al. 1997), possibly experiencing tidal
shaking or 'harrassment' (Moore, Katz, Lake 1996).  Only a small
percentage of these objects are mergers. The outer parts of the disks
are stripped during this process and/or star formation may enhance the
inner stellar densities.  Large disk--to--bulge ratios are transformed
into low disk--to--bulge ratios, i.e., early spirals may turn into S0
galaxies or, maybe, disky Es, late spirals possibly into dwarf
Es. Today, all these objects will have only modest luminosities and do
not enter the ballpark of giant Es.  At modest luminosities, however,
S0s and disky Es are indeed the dominant galaxy population in
present-day clusters (e.g., Saglia et al. 1993, J\o rgensen et
al. 1994).

As this discussion shows, the conclusions derived from the
redshift evolution of Es are in good agreement 
with the stellar population properties of local Es.

\acknowl  Thanks go to the organizers for arranging a very 
stimulating conference and for financial support. I am grateful
to my collaborators P.Belloni, L.Greggio, U.Hopp,
R.Saglia, D.Thomas and B.Ziegler for discussions,
comments and for allowing me to present work in progress.
This research was supported by the DFG with SFB 375 and by the 
Max-Planck-Gesellschaft.

\references

Aragon--Salamanca, A., Ellis, R.S., Couch, W.J., Carter, D. (1993): 
MNRAS 262, 784

Barnes, J.E. (1996): IAU Symp. 171, p.191

Barnes, J.E., Hernquist, L. (1996): ApJ 471, 115

Belloni, P., Bender, R., Hopp, U., Saglia, R.P., Ziegler, B. (1997):
in {\it HST and the High Redshift Universe}, eds. Tavir et al., World
Scientific Press

Bender, R. (1988a): A\&A 193, L7

Bender, R. (1988b): A\&A 202, L5

Bender, R. (1990): in {\it Dynamics and Interactions of Galaxies},
ed. R. Wielen, Springer Verlag, Heidelberg, p.232           

Bender, R., Surma, P., D\"obereiner, S., M\"ollenhoff, C., Madejsky,
R. (1989): A\&A 217, 35

Bender, R., Surma, P. (1992): A\&A 258, 250   

Bender, R., Burstein, D., Faber, S. (1992): ApJ 399, 462

Bender, R., Burstein, D., Faber, S. (1993): ApJ 411, 153

Bender, R., {\it et al.} (1993): in {\it Structure,Dynamics   %
and Chemical Evolution of Elliptical Galaxies}, ESO/EIPC workshop,
eds. J. Danziger et al., European Southern Observatory, M\"unchen

Bender, R., Ziegler, B., Bruzual, G. (1996): ApJ 463, L51

Bender, R., Saglia, R.P., Ziegler, B., Belloni, P., Greggio, L.,
Hopp, U., Bruzual, G. (1997), Nature submitted

Bower, R.G., Lucey, J.R., Ellis, R.S. (1992): MNRAS, 254, 601

Burstein, D., Davies, R.L., Dressler, A., Faber, S.M., Lynden-Bell, D.,
Terlevich, R., Wegner, G. (1988): in {\it Towards Understanding Galaxies at
Large Redshift}, eds. R.G. Kron \& A. Renzini, p.17, (Kluwer, Dordrecht)

Butcher, H., Oemler, G. (1978): ApJ 219, 18

Ciotti, L., Lanzoni, B., Renzini, A. (1996): MNRAS 282, 1

Dickinson, M. (1995): ASP Conf. Ser. 86, 283

Djorgovski, S., Davis, M. (1987): ApJ 313, 59

Djorgovski, S., de Carvalho, R.R., Han, M.S. (1989): ASP. Conf. Ser. 4, 329

Dressler, A., Gunn, J. (1983): ApJ 270, 7

Dressler, A., Lynden-Bell, D., Burstein, D., Davies, R.L., Faber, S.M.,
Terlevich, R.J. \& Wegner, G. (1987): ApJ 313, 42

Dressler, A., Oemler, A., Sparks, W.B., Lucas, R.A. (1994): ApJ 435, L23

van Dokkum, P., Franx, M. (1996), MNRAS 281, 985

Dubinski, J. (1994): ApJ 431, 617

Faber, S.M., Tremaine, S., Ajhar, E., Byun, Y.-I., Dressler, A., Gebhardt, K.,
Grillmair, C., Kormendy, J., Lauer, T., Richstone, D. (1997): AJ, in press

Faber, S.M., Trager, S., Gonzalez, J., Worthey, G. (1995):
in {\it Stellar Populations}, IAU Symp. 164, eds. P.C. 
van der Kruit \& G. Gilmore, Kluwer Dordrecht      

Fisher, D., Franx, M., Illingworth, G. (1996): ApJ 448, 119

Franx, M., Illingworth, G. (1988): ApJ 327, L55

Franx, M. (1993): PASP 105, 1058 

Franx, M. (1995): IAU Symp. 164, p. 269

Frenk, C., White, S.D.M., Davis, M., Efstathiou, G. (1988): ApJ 327, 507 

Fuhrmann, K., Axer, M., Gehren, T. (1995): A\&A 301, 492

Glazebrook, K., Peacock, J., Miller, L., Collins, C. (1995): MNRAS 275, 169

Gonzalez, J.J. (1993): PhD Thesis, University of Santa Cruz

Graham, A., Colless, M. (1997), MNRAS in press

Greggio, L. (1997): MNRAS, in press

Hernquist, L., Barnes, J.E. (1991): Nature 354, 210

Jedrzejewski, R.I., Schechter, P.L. (1988): ApJ 330, L87

de Jong, R., Davies, R.L. (1996),  IAU Symp. 171, p. 358

J\o rgensen, I., Franx, M., Kj\ae rgaard P. (1994): ApJ 433, 533

J\o rgensen, I., Franx, M., Illingworth, G.D. (1996): MNRAS 280, 167

J\o rgensen, I. (1997), MNRAS, in press

Katz, N. (1992): ApJ 391, 502

Kauffmann, G., Guiderdoni, B., White, S.D.M. (1994), MNRAS 274, 153

Kauffmann, G. (1996): MNRAS in press

Kauffmann, G., Charlot, White, S.D.M. (1997): MNRAS submitted

Kormendy, J., Bender, R. (1996): ApJ 464, L119

Kormendy, J., Sanders, D.B. (1992): ApJ 390, L53

Lilly, S., Le Fevre, O., Hammer, F., Crampton, D., Schade, D.J.,

Lucey, J., Guzm\'an, R. (1993): ESO/EIPC Workshop ``Structure,
Dynamics and Chemical Evolution of Elliptical Galaxies'', Eds

Matteucci, F., Greggio, L. (1986): A\&A 154, 279

Merritt, D. (1997), this conference

Moore, B., Katz, N., Lake, G. (1996): IAU Symp. 171, p. 203

Nieto, J.-L. (1988): Bol. Acad. Nac. Cien. Cordoba, 58, 239

Nieto, J.-L., Bender, R., Surma, P. (1991): A\&A 244, L37

Pahre, M.A., Djorgovski, S.G., de Calvalho, R.R. (1996): ApJ 456, L79

Paquet, A. (1994): PhD thesis, University of Heidelberg  

Renzini, A., Ciotti, (1993): ApJ 223, 707

Rix, H.--W., White, S.D.M. (1990): ApJ 362, 52

Saglia, R.P., Bender, R., Dressler, A. (1993): A\&A 279, 75

Saglia, R.P., Bender, R., Ziegler, B., Belloni, P., Greggio, L., Hopp,
U. (1997), in preparation

Schade, D., Carlberg, R., Yee, H.K.C., Lopez-Cruz, O., Ellingson,
E. (1996): ApJ 464, L63

Schweizer, F. (1990): in {\it Dynamics and Interactions of Galaxies},
Ed. R. Wielen, Springer Verlag Heidelberg, p. 232

Schweizer, F., Seitzer, P., Faber, S.M., Burstein, D., Dalle Ore,
C.M., Gonzalez, J.J. (1990): ApJ 364, L33

Scorza, C., Bender, R. (1995): A\&A 293, 20

Stanford, S.A., Eisenhardt, P.R.M., Dickinson, M. (1995): ApJ 450, 512

Statler, T.S. (1994): ApJ 425, 500

Steinmetz, M., M\"uller, E. (1994), A\&A 281, L97

Steinmetz, M. (1995): in {\it Galaxies in the Young  
Universe}, ed. H. Hippelein, Springer Verlag, Heidelberg

Thomas, D., Greggio, L., Bender, R. (1997), in preparation

Truran, J., Thielemann, F. (1986): in {\it Stellar Populations}, eds.
C. Norman et al. Cambridge University Press      

Truran, J., Burkert, A. (1995): in {it Panchromatic
View of Galaxies}, eds. G Hensler et al., Editions Frontieres,
Gif-sur-Yvette, p.389              

Wirth, G., Koo, D., Kron, R. (1994): ApJ 435, L105

White, S.D.M. (1995): Formation and Evolution of Galaxies, Les Houches

Worthey, G., Faber, S.M., Gonzalez, J.J. (1992): ApJ 398, 69

Worthey, G. (1994): ApJS 95, 107

Worthey, G. (1996): ASP Conf. Ser. 98, 467

de Zeeuw, T., Franx, M. (1991), ARAA 29, 239

Ziegler, B., Bender, R. (1997), in preparation

\bye